\documentstyle[aps,twocolumn]{revtex}

\begin{document}
\draft
\title{Evidence of a $T=0$ Quantum Critical Point Associated with the Crossover
from Weak to Strong Localization}
\author{A. M. Mack, N. Markovi\'{c}, C. Christiansen, G. Martinez-Arizala, and A. M.
Goldman}
\address{School of Physics and Astronomy, University of Minnesota, \\
Minneapolis, MN 55455, USA}
\date{\today}
\maketitle

\begin{abstract}
A crossover between logarithmic and exponential temperature dependence of
the conductance (weak and strong localization) has been observed in
ultrathin films of metals deposited onto substrates held at liquid helium
temperatures. The resistance at the crossover is well defined by the onset
of a nearly linear dependence of conductance on thickness at fixed
temperature in a sequence of {\it in situ} evaporated films. The results of
a finite size scaling analysis treating thickness as a control parameter
suggest the existence of a $T=0$ quantum critical point which we suggest is
a charge, or electron glass melting transition.
\end{abstract}

\pacs{PACS numbers: 71.30.+h, 05.70.Jk, 72.15.Rn, 73.50.-h }




Recently there has been considerable interest in electrical transport in two
dimensional (2D) conductors. Since the work of the so-called ''gang of
four'' it has been the conventional wisdom that in two dimensions weak
disorder will localize electrons making it impossible for true metallic
behavior to occur \cite{GangFour}. Although the classical work focused on
the problem of noninteracting electrons, it has been widely accepted that
its conclusions were applicable to interacting systems as well. This view
has recently been strongly challenged by experimental studies of ultrahigh
mobility low carrier density Si MOSFETs \cite
{Kravchenko1,Kravchenko2,Kravchenko3,Popovic} which have revealed in an
elegant way a metal-insulator transition (MIT) in a two dimensional system.
Dobrosavljevi\'{c} {\it et~al}.\ \cite{Dobrosavljevic} in light of the work
of Refs.~\cite{Kravchenko1,Kravchenko2,Kravchenko3} recently reconsidered
the scaling theory of the MIT in two dimensional interacting systems and
asserted that the existence of a 2D metal-insulator transition does not
contradict any general principles. Another recent development in the physics
of two dimensional conductors was the observation of Hsu {\it et~al}.\ \cite
{Hsu} of a change in the sign of the magnetoconductance of ultrathin films
with strong spin-orbit interactions. This was correlated with a change in
the temperature dependence of the conductance from logarithmic to
exponential with decreasing temperature, {\it i.\ e.}\ the crossover between
weak and strong localization, an effect which had been observed previously%
\cite{Liu,Lee}.

Here we present the result of a detailed study of the dependence of the
conductance on thickness ($d$) and temperature in ultrathin Pd \cite{Liu}
and Bi films, metals with strong spin-orbit interactions. The success of a
finite-size scaling analysis of these data leads us to suggest that there is
a quantum critical point (QCP) in these films with the Drude conductance of
the film as the control parameter. A QCP is found at zero temperature when a
ground state is controlled by an external parameter \cite{Sondhi}. It is our
suggestion that the QCP\ we have found may be a charge or electron glass
melting transition. This is motivated by the possibility that features of
the description of vortex hopping in a two dimensional disordered
superconductor may be similar to electron hopping in a two dimensional
disordered conductor\cite{Fisher and Young}. Although the vortex-glass phase
transition in two dimensions is suppressed to $T=0$, a 2D vortex glass
correlation length which diverges in the limit of zero temperature has been
reported\cite{Dekker}. This implies that there may be a QCP in this system
as well. Finally there is ample evidence that high resistance films of
metals whose conductance is governed by hopping exhibit glass-like
properties, whereas low resistance films whose conductance is logarithmic in
temperature, and which are not superconducting are not glass-like\cite
{Martinez-Arizala}.

Our application of finite-size scaling to the weak-to-strong localization
crossover was motivated by the observation that although the crossover
appears gradual as a function of temperature as in Ref.\cite{Hsu}, there is
actually a very well-defined crossover conductance, $G_{cr}$, which is
particularly sharp especially when the thickness,\ $d,$ is the independent
variable. We find $G_{cr}\sim (26\pm 3k\Omega )^{-1}$ and $G_{cr}\sim (56\pm
6k\Omega )^{-1}$ for Bi and Pd films respectively. Although $G_{cr}$ is
different in these materials, the critical exponents of the scaling analysis
are nearly identical within experimental uncertainty.

Film thicknesses ranged from 8 to 20\AA . Bismuth films were grown on a
9\AA\ pre-deposited layer of amorphous Ge ({\it a}-Ge), whereas Pd films
were grown directly on glazed ceramic substrates \cite{Liu}. Depositions
were carried out {\it in situ} under UHV conditions ($<10^{-9}$ Torr) with
substrate temperatures held below 15K. The ultra-high vacuum environment of
the samples was sustained over an extended period so that film thickness
could be built up by a series of sequential depositions without
contamination. For a given set of films the deposition conditions
(evaporation rate, source and substrate temperature, and background
pressure) were fixed to avoid changes in morphology from film-to-film in a
sequence. Resistance was measured between deposition increments using
standard four-probe techniques with films current-biased at values less than
1nA. Current-voltage characteristics were linear up to currents at least ten
times this value.

It is generally accepted that films of metals deposited on {\it a}-Ge
substrates held at helium temperatures are continuous and homogeneous for
two reasons: first, measurable conductance is often found at the order of
monolayer coverages. Second, at very early stages of growth, the conductance
becomes a linear function of thickness. Palladium films are known to wet
glass substrates, also yielding films with measurable conductances at
monolayer coverages. Close examination of our data indicates that a linear
dependence of the conductance on thickness is never really found for films
at the very earliest stages of their growth, even somewhat beyond monolayer
coverage. An analysis of the onset of nearly linear behavior with increasing
thickness is the subject of this work.

Figure \ref{G vs d} shows the conductance, $G$, of a Bi film as a function
of $d$ at several representative temperatures, and the derivative of $G(d)$
with respect to $d$ at 3.3K. These data are obtained by sorting data on $%
G(T) $ obtained from films of different thicknesses, and considering them as
a function of $d$ at fixed temperature. The derivative is then computed
numerically. Graphs of the derivative, of which only one example is shown,
exhibit a pronounced change at a temperature dependent thickness $d_{c}$
which always occurs at a constant conductance $G_{cr}$. Films with
thicknesses greater than $d_{c}$ exhibit a nearly linear dependence on
thickness. The values of the critical conductance for Bi and Pd films, $%
G_{cr}\sim (26\pm 3k\Omega )^{-1}$, and $(56\pm 6k\Omega )^{-1}$,
respectively, are very close to $e^{2}/h$ and $e^{2}/2h$.

Explanations of $G_{cr}$ being a consequence of a structural change
occurring at a particular thickness can be ruled out. If such were the case,
the boundary between the linear and non-linear dependences would occur at a
fixed thickness independent of temperature rather than at a fixed
conductance $G_{cr}$. It should also be mentioned that the conductivity near
treshold is not described by a percolation model with the conductance given
by a power law in $(d-d_{p}),$ where $d_{p}$ is the thickness at the
threshold for conductance. In this approach the average thickness is assumed
to be proportional to the surface coverage \cite{Stauffer}. It is actually
given by an exponential in $1/d^{\alpha },$ although the power $\alpha $ is
somewhat uncertain because of the limited range of the data. A value of $%
\alpha =1/2$ would be consistent with variable range hopping \cite{ES}.

Further support of the view that the crossover is an intrinsic effect comes
from the temperature dependence of the conductance at fixed thickness, which
is not shown. At low temperatures, and in the thinnest films at all
temperatures, the conductance is activated. For resistances below $%
50-100k\Omega $ it is given by the well-known functional form of the soft
Coulomb gap model $G\propto \exp [-(T_{0}/T)^{1/2}]$\cite{ES}. Here $T_{0}$
is related to the localization length, $L_{Loc}$, through $%
k_{B}T_{0}=e^{2}/\kappa L_{Loc}$, where $\kappa $ is the dielectric
constant. For resistances above 50-100k$\Omega $ the exponent is closer to
4/5 rather than 1/2 which may be evidence of collective variable range
hopping of charges, or may be the result of a crossover from variable range
to fixed range hopping with increasing resistance. At high temperatures for
all but the most resistive films the conductance varies linearly with $\ln T$
consistent with the expectations of the electron interaction picture\cite
{Altschuler}. Because the crossover in temperature between logarithmic and
activated conductance is not as sharp as the thickness-dependent crossover,
it has not been used to determine $G_{cr}$. The fact that the crossover is
seen at the same $G_{cr}$ in both the temperature and thickness dependence
of the conductance suggests that the effect is not some manifestation of
percolation on an atomic scale\cite{perc} . We have also observed a change
in the sign of the magnetoresistance in Bi films from negative to positive
with increasing thickness which is qualitatively consistent with Ref.\cite
{Hsu}. However, it does not occur at the same value of the resistance as the
crossovers described above. This delicate issue is currently under study.

The above observations led us to investigate the possibility of a quantum
critical point (QCP). In this scenario for $T>0$ there would be a line of
crossovers at $d=d_{c}(T)$, terminating in a critical point at $T=0$. It is
sufficient for a QCP that the finite temperature boundary be a crossover. It
is not necessary that it be a line of first order transitions \cite
{Abrahams-Kotliar}. Because we can determine $G_{cr}$ from $G(d)$ with
precision we take thickness $d$ as the control parameter, and define $d_{c}$
as a temperature dependent critical thickness. The use of a
temperature-dependent parameter will be seen to be a convenience, with the
Drude conductance at high temperatures, which is temperature independent,
being the actual control parameter.

The distance from the critical thickness, $\delta =d-d_{c}$ is then the
relevant field. Standard arguments \cite{Sondhi,Abrahams-Kotliar,Grupp}
suggest that the scaling form for a two dimensional system is 
\begin{equation}
G(\delta ,T)=G_{cr}F(\delta T^{-1/\nu z})  \label{one}
\end{equation}
Here $F(x)$ is a universal scaling function such that $F(0)=1$, $\nu $ is
the coherence length exponent, and $z$ is the dynamical critical exponent.
To proceed with the analysis we rewrite Eq.\ \ref{one} 
\begin{equation}
G(\delta ,t)=G_{cr}F(\delta t)
\end{equation}
and define $t\equiv T^{-1/\nu z}$. The parameter $t(T)$ is treated as an
unknown variable to be determined at each temperature to obtain the best
collapse of all the data \cite{Grupp}. The exponent $\nu z$ is then found
from the temperature dependence of $t$, which must be a power law in
temperature for the procedure to make physical sense. We should note that
the present scaling procedure does not require detailed knowledge of the
functional form of the temperature dependence of the conductivity, or prior
knowledge of the critical exponents. It is simply based on the data which
includes an independent determination of $d_{cr}$ at each temperature.

The results of the analysis for both Bi and Pd are shown together in Fig.~%
\ref{Collapse}. The crossover thickness $d_{c}(T)$, determined from the data
like that of Fig.~\ref{G vs d}, increases logarithmically with decreasing
temperature, as shown on Fig. \ref{dc vs T}. The temperature dependence of
the parameter $t,$ shown on Fig. \ref{T vs t} is consistent with a power
law, with $\nu z=6.9\pm 0.7$ and $7.2\pm 1.0$ for Bi and Pd, respectively.

Although our analysis is carried out for convenience with an apparently
temperature-dependent critical parameter, it can be cast in terms of a
temperature independent one. Adding metal sequentially to a quench-condensed
film controls disorder through its smoothing of the random potential. What
probably also occurs is an increase in the carrier concentration, and thus
an increase in the screening. These effects are measured directly by the
conductance, and indirectly by the thickness. Although $G$ is always a
temperature dependent quantity its value at high temperatures is close to
the Drude conductance $G_{D},$ which is proportional to the thickness and
constant for a given film. A natural control parameter for the QCP\ would be 
$G_{D}$, with the relevant field being $\delta _{D}=|G_{D}-G_{cr}|.$ Because
all the films we used in the scaling analysis exhibit lnT behavior in their
conductances, with $G\propto d$ at high temperatures, the control parameter
used in our analysis, $\delta ,$ is effectively the same as $\delta _{D}$
provided that $d_{c}(T)$ is a logarithmic function of temperature as is
demonstrated in Fig.\ref{dc vs T}. In other words, temperature dependence of 
$d_{c}$ is not a consequence of an underlying finite temperature phase
transition, it only reflects the fact that $d_{c}$ is related to $G_{cr}$
through a temperature dependent factor.

The crossover will always occur in films which are effectively two
dimensional at nonzero temperature because the inelastic scattering length, $%
l_{in}$ , will always be greater than the film thickness. One can crudely
estimate $l_{in}$ by equating it to the localization length at the crossover
conductance \cite{Lee-Rama}. This is tantamount to matching the length
scales of weak and strong localization at the crossover. Taking the
dielectric constant to be that of Ge, $\kappa =16$, one finds $l_{in}$ $\sim
100\AA $ in the thinnest films at the crossover. The inelastic scattering
length, $l_{in},$ is a power law in temperature proportional to $T^{-1/2}$
which would increase from $\sim 100\AA $ with decreasing temperature. Since $%
d_{c}$ increases as $\propto \ln T$ from $\sim 12\AA $ over the measured
temperatures, one would expect $d_{c}$ to remain smaller than the
localization length at all nonzero temperatures. Examination of the data of
Fig. 2 suggests that the temperature dependence of $d_{c}$may be saturating
at the low end of the temperature range, which may mean that $d_{c}$does not
increase indefinitely and is bounded at temperatures below the measured
range.

In previous work done by our group \cite{Liu} and others \cite{Hsu}, the
data for the thickness and temperature dependences of sequences of films
have been scaled by adjusting only the temperature axis. This procedure did
not include a specific recognition of the crossover phenomena shown in Fig.\ 
\ref{G vs d}, and involved forcing a functional form on high temperature
data which was slightly different from that of localization theory \cite
{GangFour}.

The QCP separates films whose conductances exhibit logarithmic from those
with exponential dependence on temperature at nonzero temperatures. Although
both types of films are nominally insulating, the analysis would suggest
that they have either fundamentally different ground states or perhaps
different dynamical behavior. The existence of a serious distinction between
insulators exhibiting logarithmic and exponential dependence on temperature
was first noted in the work of Burns and Chaikin \cite{Chaikin} on
thermopower in Pd and Pd-Au films. They found a sharp crossover in the
thermopower from metallic-like to insulating-like behavior at a resistance
near $h/e^{2}$. There is another important difference between films whose
conductances exhibit a logarithmic dependence on temperature and those with
an exponential dependence. The former appear {\it never} to be glass-like in
their behavior, whereas the latter are unequivocally glass-like\cite
{Martinez-Arizala}. The evidence for this is the observation of long
relaxation times and memory effects in a capacitive charging experiment.
Glass-like behavior disappeared above a characteristic temperature roughly
corresponding to the crossover from hopping to a logarithmic dependence of
the conductance on temperature.

The magnitude of the critical exponent products we have found ($\upsilon
z\sim 7$) seem large. In the absence of a detailed theory, it is hard to
comment on this. However the value is not very different from the exponent
products $\nu z$ ranging between 4 and 10 found by Dekker {\it et al.}\cite
{Dekker} in their study of the vortex-glass phase transition suppressed to
zero temperature in a 2D superconductor. Thus, the large values of critical
exponents in the present work would be plausible if charges in disordered
ultrathin films and vortices in two dimensional superconductors behaved
similarly. This line of argument leads us to {\it suggest} that the QCP
inferred from the analysis we have presented may be the melting of a charge
or an electron glass, with the higher-resistance phase being glass-like with
activated transport, and the lower resistance phase being a liquid-like with
logarithmic corrections to the classical conductivity from electron-electron
interaction effects.

We gratefully acknowledge numerous useful discussions with Myron Strongin,
Daniel Grupp, Boris Shklovskii, Anatoly Larkin and Subir Sachdev. This work
was supported in part by the National Science Foundation under Grant No.\
NSF/DMR-9623477.

\begin{figure}[tbp]
\caption{$G$ vs.\ $d$ for Bi films at 0.7K ($\circ $), 1.5K ($\diamond $),
3.3K ($\bullet $), 7.0K ($\bigtriangleup $) and 15.0K ($\bigtriangledown $).
A fit to the linear part of the 3.3K curve is shown. Also plotted is the
derivative of the 3.3K curve ($\times $). The horizontal dashed line
represents the slope of the linear fit. A sharp bend in the derivative is
observed very close to the thickness at which $G(d)=e^{2}/h$. Efforts to fit
these curves to a percolation model, $G\propto (d-d_{p})^{\gamma }$ were
unsuccessful. }
\label{G vs d}
\end{figure}

\begin{figure}[tbp]
\caption{Bi ($\circ$) and Pd ($\bullet$) data collapsed together. The poor
collapse of the Pd data in the lower right is from the highest temperature
curves ($T \geq 10K$), indicating that the scaling breaks down far from the $%
T=0$ transition.}
\label{Collapse}
\end{figure}

\begin{figure}[tbp]
\caption{$d_c$ vs.\ $T$ for Bi ($\circ$) and Pd ($\bullet$). $d_c$ increases
slowly (logarithmically) with decreasing temperature.}
\label{dc vs T}
\end{figure}

\begin{figure}[tbp]
\caption{$T$ vs.\ $t(T)$ for Bi ($\circ$) and Pd ($\bullet$). The slopes of
the power law fits yield $\nu z=7.2\pm 0.5$ and $\nu z=6.9\pm 0.5$ for Bi
and Pd, respectively.}
\label{T vs t}
\end{figure}

\end{document}